%% file: main.tex
\documentclass[conference,peer-reviewed]{aesconf}

\usepackage[utf8]{inputenc} % UTF-8 encoding
\usepackage{amsmath,amssymb}

\usepackage{microtype}      % Better typography
\usepackage{graphicx}
\usepackage[cal=cm]{mathalfa} 

\usepackage{booktabs}
\usepackage{xcolor}
\usepackage{hyperref}

\usepackage{subcaption}
\usepackage{adjustbox}

\usepackage[nolist]{acronym}

\makeatletter
 \DeclareRobustCommand\ref{
    \@ifstar\@refstar\T@ref
  }
  \DeclareRobustCommand\pageref{
    \@ifstar\@pagerefstar\T@pageref
  }
 \makeatother

\begin{acronym}
\acro{stft}[STFT]{short-time Fourier transform}
\acro{fft}[FFT]{fast Fourier transform}
\acro{istft}[iSTFT]{inverse short-time Fourier transform}
\acro{dnn}[DNN]{deep neural network}
\acro{mlp}[MLP]{multi-layer perceptron}
\acro{pesq}[PESQ]{perceptual evaluation of speech quality}
\acro{psd}[PSD]{power spectral density}
\acro{rir}[RIR]{room impulse response}
\acro{snr}[SNR]{signal-to-noise ratio}
\acro{lstm}[LSTM]{long short-term memory}
\acro{polqa}[POLQA]{Perceptual Objective Listening Quality Analysis}
\acro{sdr}[SDR]{signal-to-distortion ratio}
\acro{estoi}[ESTOI]{extended short-term objective intelligibility}
\acro{tf}[T-F]{time-frequency}
\acro{lsd}[LSD]{log spectral distance}
\acro{sisdr}[SI-SDR]{scale-invariant signal to distortion ratio}
\acro{mos}[MOS]{mean opinion score}
\acro{clap}[CLAP]{Contrastive Audio Language Pretraining}
\end{acronym}

\title{Automatic Audio Equalization with 
Semantic Embeddings}

\author[1]{Eloi Moliner}
\author[1]{Vesa Välimäki}
\author[2]{Konstantinos Drossos}
\author[2]{Matti S. Hämäläinen}

\affil[1]{Acoustics Lab, Department of Information and Communications Engineering, Aalto University, Espoo, Finland}
\affil[2]{Nokia Technologies, Tampere/Espoo, Finland}

\correspondence{Eloi Moliner}{eloi.moliner@aalto.fi}

\lastnames{Moliner, Välimäki, Drossos, and Hämäläinen}

\shorttitle{Automatic Audio Equalization}

\begin{document}

\twocolumn[
\maketitle % MANDATORY!

\begin{onecolabstract}
This paper presents a data-driven approach to automatic blind equalization of audio by predicting log-mel spectral features and deriving an inverse filter. The method uses a deep neural network, where a pre-trained model provides semantic embeddings as a backbone, and only a lightweight head is trained.
This design is intended to enhance training efficiency and generalization.
Trained on both music and speech, the model is robust to noise and reverberation.
Objective evaluations confirm its effectiveness, and subjective tests show performance comparable to that of an oracle that uses true log-mel spectral features, indicating that the model accurately estimates the desired characteristics,
with remaining limitations attributed to the filtering stage.
Overall, the results highlight the potential of the method for real-world audio enhancement applications.
%An objective evaluation confirms its effectiveness, whereas a subjective test shows a performance comparable to an oracle that uses true log-mel spectral features, and is thus only limited by the equalization model. 
%Our whole evaluation demonstrates that the proposed method has potential for real-world applications.
\end{onecolabstract}
]

\section{Introduction}

Automatic blind equalization (EQ) of speech and music is a key aspect
%has become an important area
in audio processing. The goal is to adjust the spectrum of an audio signal to achieve a desired tonal balance without knowledge of the original recording conditions or the target equalization curve \cite{valimaki2016all}.
Such processing enhances the audio listening experience by improving clarity and consistency.
%, often using parametric equalizers
%Such processing enables and enhances data-driven automatic equalization, enhancing the audio listening quality and experience.
%~\cite{martinez2018end}.
By enabling data-driven adjustments, EQ can significantly improve sound quality across a wide range of contexts.
Automatic equalization is used in music production \cite{de2017ten, martinez2022automatic}, hearing aids \cite{sankowsky2015individual}, and teleconferencing \cite{liu2022automatic}.
%Applications of automatic equalization range audio mixing and mastering \cite{de2017ten, martinez2022automatic}, hearing aids \cite{sankowsky2015individual}  or teleconferencing \cite{liu2022automatic}.

%A related area is automatic music mixing, where the goal is to automate
%a recording mixing process, aiming to create well-balanced and pleasant-sounding mixes \cite{de2017ten, martinez2022automatic}.
%
% Applications are...
%

This problem has been studied extensively.
A well-known approach involves estimating an equalization curve by averaging the spectral characteristics of a reference recording or collection of recordings \cite{stockham1975blind, ma2013implementation}.
More recently, deep learning techniques have been explored, employing neural networks to model complex example-dependent spectral transformations. End-to-end convolutional networks have been applied to automatic EQ \cite{martinez2018end}, while methods based on self-supervised frameworks have been proposed for blind parameter estimation \cite{steinmetz2022style, peladeau2024blind, steinmetz2024stitocontrollingaudioeffects}. Closer to this work, Mockenhaupt et al. \cite{mockenhaupt2024automatic} estimated EQ targets using a classification-based approach, though the reliance on manually defined targets limits flexibility. 
%Liu et al. \cite{liu2024automatic} addressed equalization for clarity enhancement in teleconferencing and multi-track systems. 
Generative models have also been explored, with Moliner et al. applying a diffusion-based model for equalization and audio restoration \cite{moliner2024diffusion}.

While deep learning has shown promise for EQ, practical deployment requires models that generalize across diverse signals and remain robust to noise and reverberation. This paper presents a data-driven approach to blind equalization by predicting log-mel spectral features and deriving an inverse filter, explicitly optimized for robustness and generalization. 

This paper is organized as follows. Sec.~\ref{sec:problem_definition} defines the blind EQ problem as an inverse filtering task. Sec.~\ref{sec:methods} details the proposed method, which combines a pretrained feature extractor with a lightweight trainable head for parameter estimation. Sec.~\ref{sec:experiments} describes the experimentals, including training on speech and music. Secs.~\ref{sec:objective} and~\ref{sec:subjective} present objective and subjective evaluations. Finally, Sec.~\ref{sec:conclusions} summarizes key findings.

\section{Problem Definition}
\label{sec:problem_definition}

\input{problem_definition}
\section{Methods}
\label{sec:methods}

\input{methods}

\section{Experiments}
\label{sec:experiments}
\input{experiments}

\section{Objective Evaluation}

\label{sec:objective}

\input{objective_results}

\section{Subjective Evaluation}
\label{sec:subjective}

\input{subjective_eval}

\section{Conclusions}
\label{sec:conclusions}

This paper introduces a method for automatic audio equalization using blind power spectrum estimation combined with inverse filtering, showcasing reliable performance across various challenging conditions. The approach has been designed to work with both speech and music and to be robust to noise and reverberation. 
The proposed method shows a distinct advantage when processing audio with varied semantic content, such as music, while its benefit over baseline methods is less pronounced for speech.
Our listening test results indicate a marked preference for this method in different music genres, achieving results close to Oracle level without the need for reference data.
However, the gap between the proposed method and the Reference baseline highlights opportunities for further improvement.
Current limitations are primarily linked to the inverse filtering stage, particularly the use of a subsampled spectrum via a mel-scale filterbank.

\section{Acknowledgment}

This work has been
financed in part by Nokia Technologies through the DeepMix project
(Aalto University project no. 411116).

\bibliographystyle{unsrt}

% Reference to bibliography file.
\bibliography{refs}

\end{document}

%% file: problem_definition.tex
%Let $\{ \mathbf{x}_i \}_{i=0}^{N-1}$ be a set of unknown sources, where each $\mathbf{x}_i \in \mathbb{R}^L$ represents an audio waveform. 
%We consider the observed signals $\{ \mathbf{y}_i \}_{i=0}^{N-1}$, with each $\mathbf{y}_i \in \mathbb{R}^L$, which are generated by the following forward model:
%\begin{equation}\label{eq:multisource}
%   \left\{ \mathbf{y}_i = \mathcal{A}_i(\mathbf{x}_i) + \mathbf{n}_i
%   \right\}_{i=0}^{N-1},
%\end{equation}
%where $\mathcal{A}_i(\cdot): \mathbb{R}^L \rightarrow \mathbb{R}^L$ is a forward operator that represents a specific degradation function, such as filtering or nonlinear transformation. We consider that different operators are applied to each source signal $\mathbf{x}_i$, and all of them are unknown. 

 %The set $\{ \mathbf{n}_i \}_{i=0}^{N-1}$  represents independent noise sources, which may arise from various issues such as environmental factors, interference, or limitations of the recording equipment.

The problem is how to automatically equalize the tonal balance of a degraded audio signal towards a reference. Let $\mathbf{x} \in \mathbb{R}^L$ represent a high-quality reference audio signal. We observe a degraded version of this signal, $\mathbf{y} \in \mathbb{R}^L$, generated through the forward model:
\begin{equation}\label{eq:forward_model}
  \mathbf{y} = \mathcal{A}(\mathbf{x}) + \mathbf{n},
\end{equation}
where $\mathcal{A}(\cdot): \mathbb{R}^L \rightarrow \mathbb{R}^L$ is a degradation operator (e.g., filtering or nonlinear distortion), and $\mathbf{n} \in \mathbb{R}^L$ captures additive noise from environmental interference, equipment limitations, or other sources. Our objective is to estimate a reconstruction $\hat{\mathbf{x}} \approx \mathbf{x}$.
%In this work, we restrict $\mathcal{A}(\cdot)$ to linear time-invariant systems. Under this assumption, 
Assuming that $\mathcal{A}(\cdot)$ is a linear time-invariant system, the degradation can be modeled as a convolution with a finite impulse response filter $\mathbf{h} \in \mathbb{R}^K$ of length $K$:
\begin{equation}\label{eq:lti_model}
  \mathcal{A}(\mathbf{x}) \approx \mathbf{h} \ast \mathbf{x},
\end{equation}
where $\ast$ denotes discrete convolution.
We further assume that the problem is well-conditioned, implying that the filter is invertible, with inverse \(\mathbf{h}^{-1}\).
In the noiseless case, the reference can be approximated as 
\begin{equation}\label{eq:time_domain_approx}
\mathbf{x} \approx \hat{\mathbf{x}} = \hat{\mathbf{h}}^{-1} \ast \mathbf{y}.
\end{equation}
Thus, the task of estimating the reference is reduced to estimating the inverse filter \(\hat{\mathbf{h}}^{-1}\). However, this approximation does not hold in the presence of noise \(\mathbf{n}\). As we will explain later, this issue is addressed by pre-processing \(\mathbf{y}\) with a denoiser.
In cases where the problem is ill-conditioned, such as when portions of the signal (e.g., certain frequency bands or segments) are lost or severely degraded, our method is limited to restoring only the nondegraded parts of the signal.

%% file: methods.tex
\subsection{Approximating the Inverse Filter}\label{sec:inverse_filter}

This work aims to estimate the inverse filter \( \mathbf{h}_i^{-1} \) without any prior knowledge of the forward filter \( \mathbf{h}_i \), making it a blind estimation task. The approach is performed in the frequency domain, where we estimate the inverse filter \( \hat{\mathbf{H}}^{-1} \in \mathbb{C}^{N_\text{FFT}} \), based on a chosen FFT size \( N_\text{FFT} \). 

Equation \ref{eq:time_domain_approx} can be written in the frequency domain as 
\begin{equation}
\mathbf{x} \approx \mathcal{F}^{-1} 
\left( 
\hat{\mathbf{H}}^{-1} \odot  \mathcal{F}(\mathbf{y})
\right),
\end{equation}
where \( \mathcal{F}(\cdot) \) and \( \mathcal{F}^{-1}(\cdot) \) denote the short-time Fourier transform (STFT) and its inverse (iSTFT), respectively. Both transformations are computed using \( N_\text{FFT} \) frequency bins, with a window size and hop size defined in Sec.\,\ref{sec:data}.
Owing to the Hermitian symmetry of the Fourier transform for real-valued signals, only 
$N_\text{FFT}/2 +1$ 
bins contain non-redundant information, and thus only these need to be estimated in practice.

We constrain the inverse filter \( \hat{\mathbf{H}}^{-1} \) to be a zero-phase filter, and approximate its magnitude as
\begin{equation}\label{eq:inverse_filter}
|\hat{\mathbf{H}}^{-1}| =  \sqrt{\frac{\mathbf{X}_\text{ref}^2}{\mathbf{Y}_\text{avg}^2}},
\end{equation}
\noindent where \( \mathbf{Y}_\text{avg}^2 \) is the time-averaged Power Spectral Density (PSD) of the observed signal \( \mathbf{y} \), computed as 
\begin{equation}\label{eq:avgspec}
\mathbf{Y}_\text{avg}^2 = \frac{1}{M} \sum_{m=0}^{M-1} |\mathrm{STFT}(\mathbf{y})_m|^2.
\end{equation}
Here, \( M \) is the number of STFT frames, and \( \mathbf{X}_\text{ref}^2 \) denotes the target or reference spectral shape, which is typically content-dependent. The reference \( \mathbf{X}_\text{ref}^2 \) could be a user-defined spectral curve, or, more generally, the average power spectrum of the unknown clean source \( \mathbf{x} \), given in an analogous way as in Eq.~\eqref{eq:avgspec}.

%\[
%\mathbf{X}_\text{ref}^2 = \frac{1}{M} \sum_{m=0}^{M-1} |\mathrm{STFT}(\mathbf{x})_m|^2
%\]

Since the clean source \( \mathbf{x} \) is unknown, we estimate the target spectrum \( \hat{\mathbf{X}}_\text{ref}^2 \approx \mathbf{X}_\text{ref}^2 \) using a deep neural network (DNN). The DNN denoted as \( E_\theta(\mathbf{y}) \) is trained to predict the target spectrum given the observed signal \( \mathbf{y} \), where \( \theta \) represents the trainable parameters of the network.

Instead of predicting the target spectral shape directly $\hat{\mathbf{X}}_\text{ref}^2$, we reduce its dimensionality by using a mel-scale triangular filterbank. The mel-transformed reference spectrum $\mathbf{z}_\text{ref} \in \mathbb{R}^{N_\text{mel}}$ is given by 
\begin{equation}
\mathbf{z}_\text{ref} = 10 \log_{10}(\mathbf{M} \mathbf{X}_\text{ref}^2),
\end{equation}
\noindent where $\mathbf{M} \in \mathbb{R}^{N_\text{mel} \times (N_\text{FFT}/2+1)}$ represents the mel-scale filterbank, which compresses the non-redundant frequency bins into fewer mel-scaled bins \( N_\text{mel} \). The DNN is trained to predict the features $\mathbf{z}_\mathrm{ref}$, such that  $E_\theta(\mathbf{y}) = \hat{\mathbf{z}} \approx \mathbf{z}_\text{ref}$.

Once we have obtained \( \hat{\mathbf{z}} \), the estimated time-averaged PSD in the original frequency domain can be computed by reversing the mel transformation:
\begin{equation}\label{eq:interp}
\hat{\mathbf{X}}_\text{ref}^2 = \mathbf{M}^T 10^{\hat{\mathbf{z}}/10}.
\end{equation}
\noindent Here, \( \mathbf{M}^T \) is the transpose of the mel filterbank, which interpolates the full FFT frequency resolution from the lower-dimensional mel representation.

\subsection{Blind Parameter Estimation}

The DNN model \(E_\theta\) is designed to estimate the log-mel spectral features \(\mathbf{z}(\mathbf{x})\) of an original recording \(\mathbf{x}\), given a processed observation \(\mathbf{y}\).
The target parameters \(\mathbf{z}(\mathbf{x})\) are extracted from the clean signal \(\mathbf{x}\) using a procedure described in Sec.~ \ref{sec:inverse_filter}. 
During training, the observation \(\mathbf{y}\) is generated by applying a forward transformation \(f(\mathbf{x})\) to the same original recording.
This transformation introduces a range of perturbations, including randomized spectral coloration, variations in loudness, room reverberation, and additive noise.
This makes the spectral parameter estimation non-trivial.

We hypothesize that \(\mathbf{z}(\mathbf{x})\), being both time-averaged and compressed through a mel filterbank, is correlated with the type of content in the input signal. For example, orchestral music tends to have a distinct spectral envelope, while speech has a different one, and characteristics like gender or age may further affect the spectral shape of speech signals. Based on this, the model \(E_\theta(\mathbf{y})\) is expected to extract semantic information from the processed signal \(\mathbf{y}\) and map it to the corresponding log-mel spectral features \(\hat{\mathbf{z}}\).

To achieve this, we propose factorizing the DNN model \(E_\theta\) into two components, such that 
\begin{equation}
E_\theta(\mathbf{y}) = (\text{MLP}_{\theta_\text{MLP}} \circ \text{CLAP}_{\theta_\text{CLAP}})(\mathbf{y}),
\end{equation}
where \(\text{CLAP}_{\theta_\text{CLAP}}\) is a pretrained audio encoder  \cite{Elizaide2024clap}, and \(\text{MLP}_{\theta_\text{MLP}}\) is a small \ac{mlp} that learns the mapping from \ac{clap} embeddings to log-mel spectral features. \ac{clap} is a joint audio-text encoder trained via contrastive learning, mapping audio and text pairs into a shared embedding space. We use the audio encoder of \ac{clap}, with its parameters frozen during training, to extract semantic features from the signal \(\mathbf{y}\).

Previous work has shown that embeddings from \ac{clap} and similar models often fail to capture the details of audio effects and transformations, such as those introduced by spectral processing, limiting their sensitivity to these changes \cite{gui2024adapting, hawley2023leveraging}.
This is beneficial in our case, as it suggests that the forward operator \(f(\cdot)\), which introduces spectral coloration, may not significantly affect CLAP-derived embeddings. Thus, \(\text{MLP}_{\theta_\text{MLP}}\) can focus on learning the relationship between the extracted embeddings and the log-mel spectral features.

The parameters of \( \mathbf{z}(\mathbf{x}) \) typically exhibit varying magnitudes, which generally decrease logarithmically as a function of frequency. Normalizing both the input and target data is beneficial during neural network training, as it ensures consistent scaling and improves model convergence. To achieve uniform error impact across all frequency bands, we first compute the average \( \mathbf{z}_\text{avg} \) coefficients from a representative subset of the training data. The model is then trained to estimate the deviation of each instance from this average, ensuring balanced attention across the different frequency components. 

The parameter estimation model \(E_\theta\) is optimized using the following loss function:
\begin{equation}\label{loss}
\mathcal{L} = \mathbb{E}_{\mathbf{x} \sim p_x, f(\cdot) \sim p_f} \left[
\lVert E_\theta(f(\mathbf{x})) - (\mathbf{z}(\mathbf{x})- \mathbf{z}_\text{avg})  \rVert_1
\right],
\end{equation}
where \(\mathbf{x}\) represents clean audio data sampled from a distribution \(p_x\) (e.g., a dataset of recordings), and the forward operator \(f(\cdot)\) is sampled from a distribution \(p_f\), representing different types of data augmentations. 
 As illustrated in Fig.~\ref{fig:pipeline}, these degradations include room reverberation, randomized spectral coloration, additive noise, and gain randomization.
Importantly, since the CLAP encoder is frozen during training, only the parameters of the MLP \(\theta_\text{MLP} \subset \theta\) are optimized based on the loss objective.
It is important to note that, due to the \(\ell_1\)-norm minimization objective, at convergence the network \(E_\theta\) approximates the conditional median of \(\mathbf{z}(\mathbf{x})\) given the transformed input \(f(\mathbf{x})\) under the training data distribution. Thus, the estimated power spectrum \(\hat{\mathbf{z}} = E_\theta(f(\mathbf{x}))\) can be interpreted as 
\begin{equation}\label{eq:median}
\hat{\mathbf{z}} \approx \operatorname{median}_{\mathbf{z} \sim p(\mathbf{z} | f(\mathbf{x}))} [\mathbf{z}-\mathbf{z}_\text{avg}] + \mathbf{z}_\text{avg},
\end{equation}
where the median is computed implicitly over the target distribution defined by the training dataset.

The training process is illustrated in Fig.~\ref{fig:training}, and the signal processing conducted at inference time is summarized in Fig.~\ref{fig:inference}.
For simplicity, the detail of $\mathbf{z}_\text{avg}$ is omitted in both Fig.~\ref{fig:training} and \ref{fig:inference}.

\input{diagram_train}

\input{diagram_inference}

%% file: diagram_train.tex
\begin{figure}[t]

\centering
\adjustbox{width=0.7\linewidth}{
\includegraphics[]{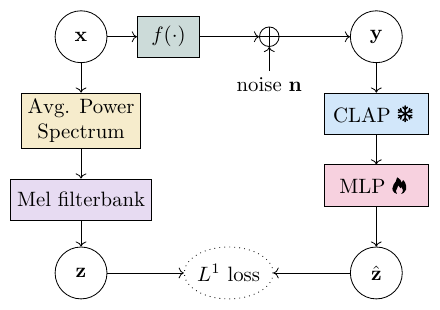}
}

\vspace{-0.2cm}
    \captionsetup{font=small}
\caption{\protect{Training diagram.}}

\label{fig:training}

\end{figure}

%% file: diagram_inference.tex
\begin{figure}[t]

\centering
\adjustbox{width=1\linewidth}{
\includegraphics[]{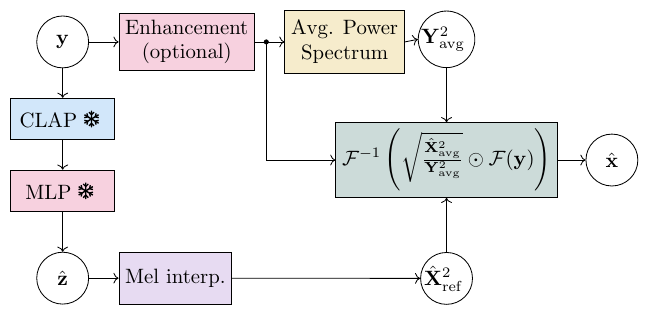}
}

\vspace{-0.2cm}
    \captionsetup{font=small}
\caption{\protect{Inference diagram.}}

\label{fig:inference}

\end{figure}

%% file: experiments.tex
\subsection{Training Data Pipeline}

We develop a single model capable of processing both speech and music. The model is designed to exhibit robustness to background noise, gain variations, and reverberation, necessitating the inclusion of these disturbances within the training data pipeline. We utilize the speech datasets VCTK \cite{valentini2016reverberant} and EARS \cite{richter2024ears}, both of which consist of high-quality studio speech recordings sampled at 48 kHz, and which overall contain around 200 different speakers.

We choose MedleyDB \cite{bittner2016medleydb} and DSD100 \cite{liutkus20172016} as the music datasets, both offering studio-quality recordings across various genres such as rock, pop, classical, and jazz. These datasets are mixed by recording engineers, allowing us to consider the spectral balance as an appropriate target. Although the datasets provide isolated stems for every source in the microphone, we limit our use to the mixed tracks.
Every audio segment undergoes loudness normalization to achieve a standardized volume level of $-18$ LUFS.
This procedure guarantees a consistent loudness target, ensuring uniformity across different audio files.

\begin{figure} [t]
    \centering
    \adjustbox{width=\linewidth}{
    \includegraphics[]{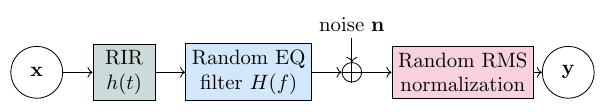}
    }
    \captionsetup{font=small}
    \caption{Pipeline illustrating the applied degradations. }
    \label{fig:pipeline}
\end{figure}

The degradation process is outlined in Fig. \ref{fig:pipeline}, illustrating the practical implementation of $f(\cdot)$ as shown in Fig. \ref{fig:training}. Initially, an audio segment with approximately 7 s in duration, containing either speech or music, is convolved with a room impulse response (RIR). We specifically use RIRs from the Arni dataset \cite{prawda2022calibrating}.
Subsequently, the audio is subjected to an equalization filter featuring random gains across various frequency bands. This filter is constructed in the frequency domain as a zero-phase FIR filter. The gain for a set of 26 bark-scaled frequency bands is stochastically sampled on a log-uniform scale ranging from -20 to 20 dB.
In the next phase, noise is introduced. For noise sources, the TAU Urban Acoustic Scenes dataset \cite{heittola_2022_6337421} is employed during training.
This dataset is integrated with speech and music by mixing with noise from this dataset, achieving an SNR that is log-uniformly sampled between 5 and 50 dB. 

Finally, the gain of the mixture is adjusted through a random root-mean-square (RMS) normalization. The RMS normalization is implemented as 
$y=(\alpha \sqrt{L}/\rVert \mathbf{x} \lVert_2) \mathbf{x}$,
with the parameter $\alpha$ being uniformly sampled within $-5$\,dB and $30$\,dB.
%the interval [0.02, 0.18].

\subsection{Training Details}

The training of the model consisted of conducting 100,000 iterations, utilizing the Adam optimization algorithm. The learning rate employed during this process was set to $1\times10^{-4}$, and the values of the beta momentum parameters were configured to be (0.9, 0.999). For the training data, audio segments of 7 s sampled at 48 kHz were used.
The training was performed with a batch size of 16.

We chose a number of mel-frequency coefficients $N_\text{mel}=32$, extracted using the STFT with FFT size 2048, a Hann window with size 1024, and a hop length of 256 samples.
The employed \ac{mlp} contains 0.34 M parameters and is based on three linear layers with a hidden size of 256 and Leaky-ReLU nonlinearities.
    
\subsection{Evaluation Data}\label{sec:data}

The model is evaluated on speech and music, and in different scenarios, according to the applied degradations. We distinguish the following scenarios: 1.~Speech equalization, 2.~Music equalization, 3.~Noisy speech equalization, and 4.~Reverberant speech equalization.
%\item Real speech recordings

For scenarios 1, 3, and 4, we evaluate on VCTK speakers ``p351'', ``p360''–``p364'', ``p374'', ``p376'' and EARS speakers ``p100''–``p107''.
%In scenarios 1, 3, and 4, we evaluate using test splits from VCTK and EARS. Specifically, we include speakers "p351", "p360", "p361", "p362", "p363", "p364", "p374", and "p376" from VCTK, along with speakers "p100", "p101", "p102", "p103", "p104", "p105", "p106", and "p107" from EARS.
In scenario 2, we extract 8-s segments from separated test splits of MedleyDB and DSD100. For each of these scenarios, a test set comprising 1,000 samples is created. To each example, a random filter is applied following the same procedure used during training, and the gain is randomized with an RMS in the interval [0.005, 0.30], extending the range seen at training time. 

In scenario 3, we introduce noise from the DEMAND dataset \cite{thiemann2013diverse} and apply SNRs ranging between $-5$\,dB and $30$\,dB, which also extends beyond the training range. Finally, in scenario 4, the signals are convolved with RIRs, utilizing simulated RIRs from pyroomacoustics \cite{scheibler2018pyroomacoustics}, with a reverberation time $T_{60}$ spanning [0.1, 1.0]. 
In both scenarios, we apply DeepFilterNet2 \cite{schroter2022deepfilternet2} to pre-process audio signals that are either noisy or reverberant. 
 DeepFilterNet2 was selected due to its open-source availability, lightweight architecture, high processing speed, and acceptable performance.
It is important to note that the original signal remains accessible to the model, while the pre-processing is integrated into the equalization pipeline, as shown in Fig. \ref{fig:inference}.

%Lastly, for scenario 5, the DAPS dataset (reference) is utilized. This dataset comprises high-quality speech recordings as well as versions reproduced using actual mobile devices in real environments. Consequently, it includes spectral coloration, reverberation, and background noise. Given the availability of the high-quality speech version, it is feasible to calculate reference-based metrics.

\subsection{Metrics}
To evaluate the performance of the model, we consider several metrics that provide a comprehensive analysis of its effectiveness in estimating the power spectrum and enhancing audio signals. These metrics capture both spectral accuracy and audio-domain fidelity.

We compute the L1 error between the target average power spectrum extracted from the reference high-quality recording and the estimated power spectrum. This metric corresponds directly to the loss function used during training, as defined in Eq.~\eqref{loss}. It provides a measure of how closely the model's predictions align with the target power spectrum, reflecting its ability to learn accurate spectral representations.

To evaluate the fidelity of the equalized audio signal, we employ the Log-Spectral Distortion (LSD), which measures differences in spectral features between the reference and predicted audio signals. The LSD is computed as 
\begin{equation}
\text{LSD} = \frac{1}{MK} \sum_{m,k} \bigl| \log(|\mathcal{F}(\mathbf{x})| + \epsilon) - \log(|\mathcal{F}(\hat{\mathbf{x}})| + \epsilon) \bigr|,
\end{equation}
where $\mathcal{F}$ represents the Short-Time Fourier Transform (STFT) operator,  $M$ is the number of time frames, $K$ is the number of frequency bins, and $\epsilon$ is a small constant (e.g., $10^{-6}$) to avoid logarithmic instabilities.

%To capture a metric closer to the training loss, we compute the Log-Mel Spectrogram Distance (LogMelDist). This measures the L1 distance between the magnitude Mel-spectrograms of the reference and predicted audio, formulated as 
%\begin{equation}
%\text{LogMelDist}(\mathbf{x}, \hat{\mathbf{x}}) = \frac{1}{M \times N_\text{mel}} \sum_{m=1}^{M} \sum_{n=1}^{N_\text{mel}} \left| \log \left( \mathbf{M} \mathcal{F}(\mathbf{x}) + \epsilon \right) - \log \left( \mathbf{M} \mathcal{F}(\hat{\mathbf{x}}) + \epsilon \right) \right|,
%\end{equation}
%where \(\mathbf{M} \in 
% \mathbb{R}^{N_\text{mel} \times N_\text{FFT}/2}\) is the Mel (triangular) filterbank operator in matrix form, and \(N_\text{mel}\) is the number of Mel frequency bins.

%Furthermore, to assess whether the model can appropriately apply the gain, we calculate what is referred to as the RMS error, which represents the power difference between the predicted signal and the reference signal:
% \begin{equation}
%     \text{RMS Error}(\mathbf{x}, \hat{\mathbf{x}}) = \left| \sqrt{\frac{1}{N}\sum_{i=0}^{N} \mathbf{x}_i^2}  -
%     \sqrt{\frac{1}{N}\sum_{i=0}^{N} \hat{\mathbf{x}}_i^2}  \right|.
% \end{equation}

While these objective metrics are useful for evaluating model performance, they are not always directly indicative of perceptual quality. For example, the equalization applied to music examples may deviate from the median power spectrum the model predicts. This is because the model's estimation inherently reflects the statistical properties of the training data, whereas real-world audio signals often exhibit unique spectral characteristics. 
Thus, while the metrics provide quantitative insights, qualitative evaluation through listening tests is essential to fully assess the method's impact on perceptual audio quality.

\subsection{Baselines}

%In this study, we examine DeepAFx \cite{steinmetz2022style}, a data-driven approach for audio effect transfer utilizing a self-supervised deep learning framework. We employ the publicly accessible code and pre-trained weights \footnote{\href{https://github.com/adobe-research/DeepAFx}{https://github.com/adobe-research/DeepAFx}}. Nonetheless, we acknowledge that, while it serves as a useful benchmark, direct comparison with the original public model is not entirely fair due to discrepancies in problem configuration, training objectives, and the variety of audio effects being modeled.
%Nonetheless, it serves as a benchmark for end-to-end learning approaches in audio processing.

To contextualize the performance of our proposed method, we designed and evaluate the proposed method against several baselines.
The first baseline, which we call  \textit{Average}, computes the average power spectrum of the training data, and uses that to compute the inverse filter in Eq.~\eqref{eq:inverse_filter}. We compute the average of the speech data and music data separately. 
By using these fixed average spectra as predictions, this baseline is meant to highlight the effectiveness of the proposed method at leveraging data-specific patterns compared to a naive global estimate.

The second baseline, which we call \textit{End-to-end}, replaces the frozen CLAP encoder with a fully trainable encoder, allowing the model to jointly optimize feature extraction and power spectrum estimation. 
We used the same neural network architecture as in the setup from Steinmetz et al. \cite{steinmetz2022style}.
This baseline aims to demonstrate the advantages of using a pre-trained representation over task-specific feature learning.

Finally, the \textit{Oracle} baseline uses parameters directly estimated from the reference high-quality audio signal, bypassing the need for a model prediction. This serves as an upper bound on performance within our framework, illustrating the best possible outcome achievable when the target parameters are perfectly known. 
%[mention limitations of this]

%% file: objective_results.tex
\subsection{Speech Equalization and Gain Adjustment}
\begin{figure*}[t]
    \centering
    \begin{subfigure}[b]{0.246\textwidth}
    \includegraphics[
    trim={0.2cm 0.4cm 0.3cm 0.3cm}, clip,
    width=\linewidth]{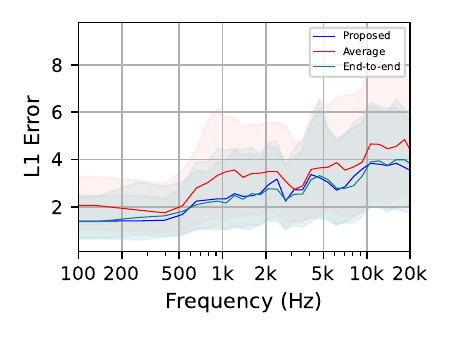}
    \vspace{-0.5cm}
    \captionsetup{font=small}
    \caption{$L_1$ error vs frequency}
    \end{subfigure}
    \begin{subfigure}[b]{0.246\textwidth}
    \includegraphics[
    trim={0.2cm 0.4cm 0.3cm 0.3cm}, clip,
    width=\linewidth]{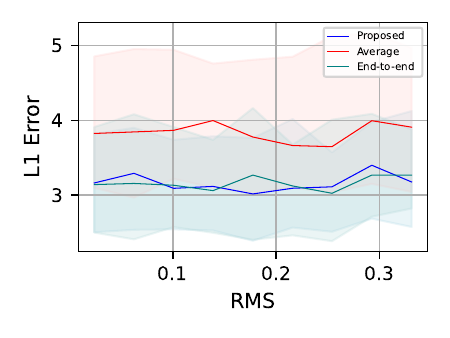}
    \vspace{-0.5cm}
    \captionsetup{font=small}
    \caption{$L_1$ error vs RMS}
    \end{subfigure}
    \begin{subfigure}[b]{0.246\textwidth}
    \includegraphics[
    trim={0.2cm 0.4cm 0.3cm 0.3cm}, clip,
    width=\linewidth]{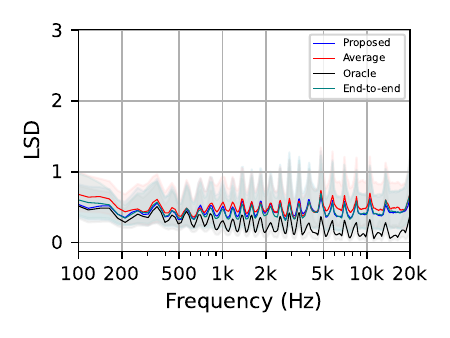} 
    \vspace{-0.5cm}
    \captionsetup{font=small}
    \caption{LSD error vs frequency}
    \end{subfigure}
    \begin{subfigure}[b]{0.246\textwidth}
    \includegraphics[
    trim={0.2cm 0.4cm 0.3cm 0.3cm}, clip,
    width=\linewidth]{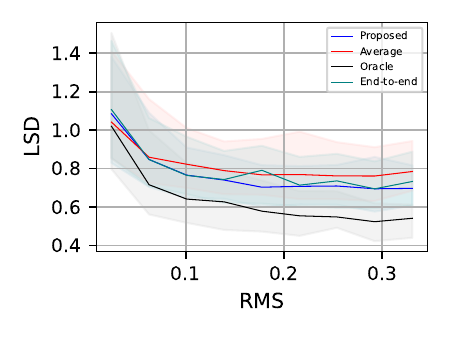}
    \vspace{-0.5cm}
    \captionsetup{font=small}
    \caption{LSD error vs RMS}
    \end{subfigure}
    %\begin{subfigure}[b]{0.32\textwidth}
    %\includegraphics[width=\linewidth]{E_speech_rms_variant_E11logmelspec_loss_rms.pdf}
    %\captionsetup{font=small}
    %\caption{LogMelDist vs RMS }
    %\end{subfigure}
    %\begin{subfigure}[b]{0.32\textwidth}
    %\includegraphics[width=\linewidth]{figures/scenario1/E_speech_rms_variant_E11rms_error_rms.pdf}
    %\captionsetup{font=small}
    %\caption{RMS Error vs Original RMS}
    %\end{subfigure}

    \vspace{-0.2cm}
    \captionsetup{font=small}
    \caption{Objective metrics from Scenario 1:  Equalization and gain adjustment for speech.}
    \vspace{-0.1cm}
    \label{fig:results1}
\end{figure*}

In this scenario, speech signals undergo arbitrary EQ and random RMS processing. The outcomes are depicted in Fig. \ref{fig:results1}.
Each plot displays the median curves represented by colored lines, with the InterQuartile Range (25\%-75\%) illustrated as shaded regions.
Figure \ref{fig:results1}a illustrates $L^1$ errors on parameter estimation in relation to frequency, highlighting increased errors in higher frequency ranges. Both the Proposed and End-to-end models display comparable performance, surpassing the Average model. Figure \ref{fig:results1}b presents $L^1$ errors concerning the RMS of the altered signal, showing consistent performance across all RMS levels, despite training only within the [0.02, 0.18] interval. Performance seems similar even beyond this range. Figure \ref{fig:results1}c depicts LSD relative to frequency, indicating that Proposed, Average, and End-to-end perform similarly, though less effectively than the Oracle condition. In Figure \ref{fig:results1}d, LSD errors in relation to RMS reveal higher errors at lower RMS values, even in the Oracle scenario, underscoring a fundamental limitation of the employed inverse filter approach (see Eq.~\eqref{eq:inverse_filter}). 
%A comparable yet milder pattern is observable in Figure \ref{fig:results1}e, which shows LogMelDist against RMS.
%Lastly, Figure \ref{fig:results1}f demonstrates RMS error versus RMS of the distorted signal, indicating uniform performance.

\subsection{Music Equalization and Gain Adjustment} \label{sec:scenario2}

\begin{figure*}[t]
    \centering
    \begin{subfigure}[b]{0.246\textwidth}
    \includegraphics[
    trim={0.2cm 0.4cm 0.3cm 0.3cm}, clip,
    width=\linewidth]{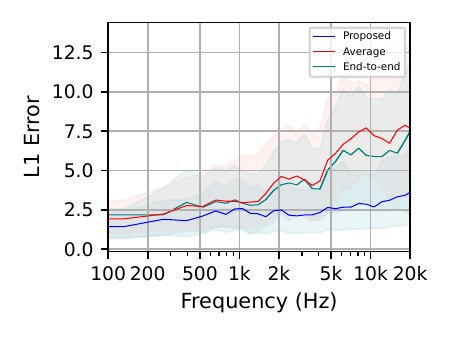}
    \vspace{-0.5cm}
    \captionsetup{font=small}
    \caption{$L_1$ error vs frequency }
    \end{subfigure}
    \begin{subfigure}[b]{0.246\textwidth}
    \includegraphics[
    trim={0.2cm 0.4cm 0.3cm 0.3cm}, clip,
    width=\linewidth]{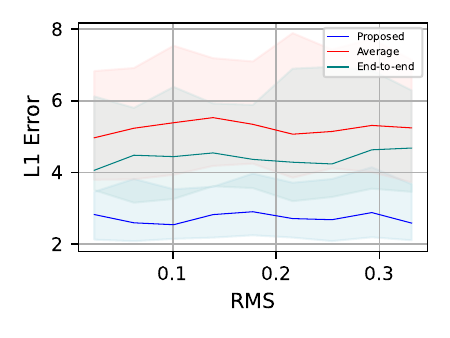}
    \vspace{-0.5cm}
    \captionsetup{font=small}
    \caption{$L_1$ error vs RMS}
    \end{subfigure}
    \begin{subfigure}[b]{0.246\textwidth}
    \includegraphics[
    trim={0.2cm 0.4cm 0.3cm 0.3cm}, clip,
    width=\linewidth]{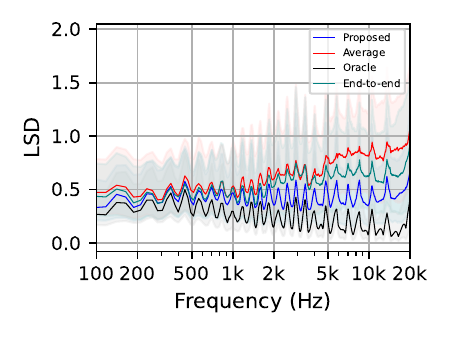} 
    \vspace{-0.5cm}
    \captionsetup{font=small}
    \caption{LSD vs frequency}
    \end{subfigure}
    \begin{subfigure}[b]{0.246\textwidth}
    \includegraphics[
    trim={0.2cm 0.4cm 0.3cm 0.3cm}, clip,
    width=\linewidth]{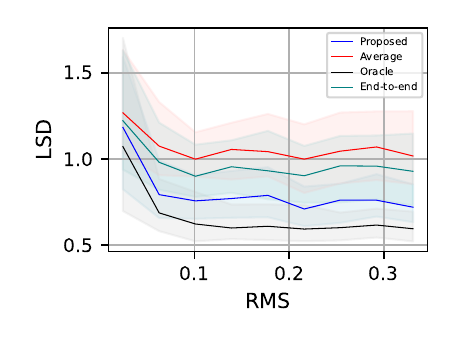}
    \vspace{-0.5cm}
    \captionsetup{font=small}
    \caption{LSD vs RMS}
    \end{subfigure}
    %\begin{subfigure}[b]{0.32\textwidth}
    %\includegraphics[width=\linewidth]{E_music_rms_variant_noiseless_E11logmelspec_loss_rms.pdf}
    %\captionsetup{font=small}
    %\caption{LogMelDist vs RMS}
    %\end{subfigure}
    %\begin{subfigure}[b]{0.32\textwidth}
    %\includegraphics[width=\linewidth]{E_music_rms_variant_noiseless_E11rms_error_rms.pdf}
    %\captionsetup{font=small}
    %\caption{RMS Error vs Original RMS}
    %\end{subfigure}
    
    \vspace{-0.2cm}
    \captionsetup{font=small}
    \caption{Objective metrics from Scenario 2: Equalization and gain adjustment for music.}
    \vspace{-0.1cm}
    \label{fig:results2}
\end{figure*}

In Scenario 2, musical signals were processed utilizing randomized equalization filters and gain levels. The outcomes, illustrated in Fig. \ref{fig:results2} and mirroring the data from Fig. \ref{fig:results1}, exhibit patterns akin to those identified in Scenario 1. Nevertheless, here, both the Average and End-to-end baselines demonstrate significant performance reduction across all metrics. This deterioration in the Average baseline’s effectiveness is probably attributable to the wider variability found in music as opposed to speech. The Proposed model sustains superior overall performance compared to the End-to-end baseline, suggesting that the use of pretrained CLAP embeddings facilitates robust feature extraction, even for the more complex music signals.
This scenario is also evaluated through subjective listening in Sec. 6.
%This Scenario is further evaluated in the subjective listening reported in Sec. 6.

\subsection{Noisy Speech Equalization}\label{sec:scenario3}

\begin{figure*}[t]
    \centering
    \begin{subfigure}[b]{0.32\textwidth}
    \includegraphics[width=\linewidth]{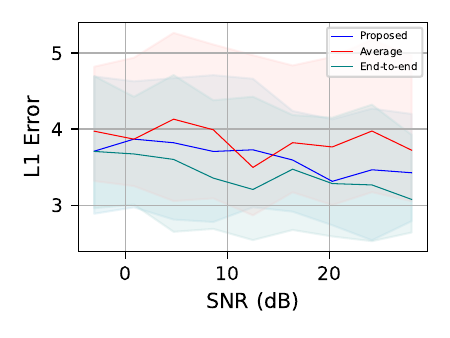}
    \vspace{-0.7cm}
    \captionsetup{font=small}
    \caption{$L_1$ error vs SNR \\ \hspace{1cm}}
    \end{subfigure}
    \begin{subfigure}[b]{0.32\textwidth}
    \centering
    \includegraphics[width=\linewidth]{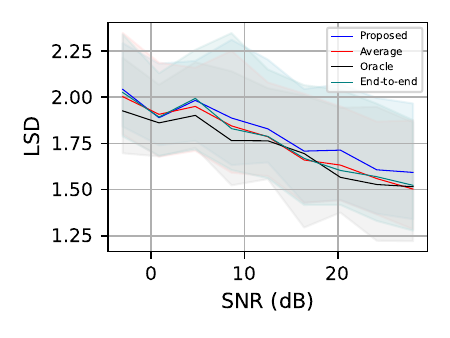} 
    \vspace{-0.7cm}
    \captionsetup{font=small}
    \centering
    \caption{LSD error vs SNR. Using speech 
 \newline enhancement.}
    \end{subfigure}
    %\begin{subfigure}[b]{0.32\textwidth}
    %\includegraphics[width=\linewidth]{E_speech_rms_variant_snr_variant_E11_less_baselineslogmelspec_loss_snr.pdf}
    %\captionsetup{font=small}
    %\caption{LogMelDist vs RMS. Using speech enhamcement.}
    %\end{subfigure}
    \begin{subfigure}[b]{0.32\textwidth}
    \includegraphics[width=\linewidth]{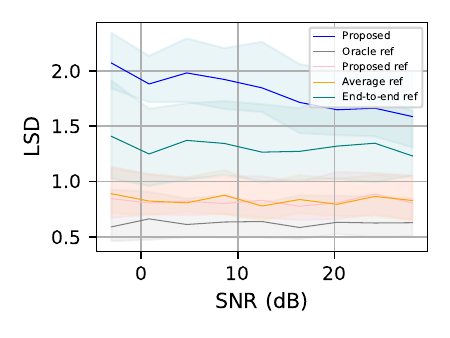} 
    \vspace{-0.7cm}
    \captionsetup{font=small}
    \centering
    \caption{LSD error vs SNR. Using denoised \newline reference.}
    \end{subfigure}
    %\begin{subfigure}[b]{0.32\textwidth}
    %\includegraphics[width=\linewidth]{E_speech_rms_variant_snr_variant_E11_reflogmelspec_loss_snr.pdf}
    %\captionsetup{font=small}
    %\caption{LSD error vs SNR. Using denoised reference.  }
    %\end{subfigure}
    \vspace{-0.2cm}
    \captionsetup{font=small}
    \caption{Objective metrics from Scenario 3: Equalization and gain adjustment for noisy speech  }
    \vspace{-0.1cm}
    \label{fig:results3}
        
\end{figure*}

This scenario investigates the impact of background noise on inverse filter estimation by analyzing the relationship between the objective metrics and the input SNR.
Figure \ref{fig:results3}a presents the $L^1$ error of the estimated parameters as a function of SNR. Both the Proposed and End-to-end methods exhibit similar performance, with a noticeable but relatively minor increase in error at lower SNRs, which aligns with expectations given the added noise.

The LSD metric, defined between audio waveforms, is affected by the speech enhancement model used for pre-processing. In Fig. \ref{fig:results3}b, we evaluate performance using DeepFilterNet2 \cite{schroter2022deepfilternet2} for speech enhancement prior to equalization. Here, all methods exhibit similar trends, heavily influenced by the SNR.
To explore an idealized scenario, we apply the estimated equalization (derived from noisy measurements) to a noiseless version of the degraded signal, simulating perfect denoising while retaining spectral coloration. As shown in Fig. \ref{fig:results3}c, this setup significantly improves LSD performance, achieving near-uniform results across the SNR range. In this case, the \textit{Proposed} and \textit{Average} conditions perform comparably, while the \textit{End-to-end} baseline shows a noticeable decline in performance.

\subsection{Reverberant Speech Equalization }

\begin{figure*}[t]
    \centering
    \begin{subfigure}[b]{0.32\textwidth}
    \includegraphics[width=\linewidth]{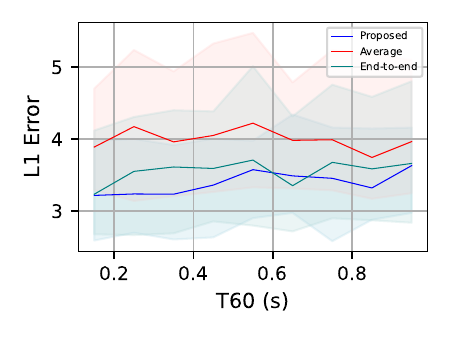}
    \vspace{-0.7cm}
    \captionsetup{font=small}
    %\caption{$L_1$ error vs reverberation time T$_{60}$ }
        \caption{$L_1$ error vs reverberation time T$_{60}$ \\ \hspace{1cm}}
    \end{subfigure}
    \begin{subfigure}[b]{0.32\textwidth}
    \includegraphics[width=\linewidth]{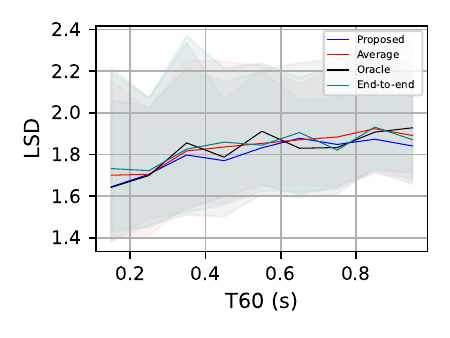} 
    \vspace{-0.7cm}
    \captionsetup{font=small}
    \caption{LSD vs reverberation time T$_{60}$. \newline Using speech enhancement.}
    \end{subfigure}
    %\begin{subfigure}[b]{0.32\textwidth}
    %\includegraphics[width=\linewidth]{E_speech_rms_variant_RIR_variant_E11_less_baselineslogmelspec_loss_t60.pdf}
    %\captionsetup{font=small}
        %\caption{LogMelDist vs reverberation time T$_{60}$. Using speech enhancement.  }
    %\end{subfigure}
    \begin{subfigure}[b]{0.32\textwidth}
    \includegraphics[width=\linewidth]{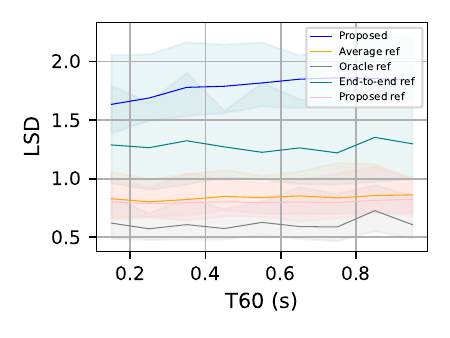}
    \vspace{-0.7cm}
\captionsetup{font=small}
    \caption{LSD vs reverberation time T$_{60}$. \newline Using dereverberated reference. }
    \end{subfigure}
    %\begin{subfigure}[b]{0.32\textwidth}
    %\includegraphics[width=\linewidth]{E_speech_rms_variant_RIR_variant_E11_reflogmelspec_loss_t60.pdf}
    %\captionsetup{font=small}
        %\caption{LogMelDist vs reverberation time T$_{60}$. Using dereverberated reference. }
    %\end{subfigure}

    \vspace{-0.2cm}
\captionsetup{font=small}
    \caption{Objective metrics from Scenario 4: Equalization and gain adjustment for reverberant speech  }
    \vspace{-0.1cm}
    \label{fig:results4}
\end{figure*}
%\subsection{Scenario 5}

This scenario builds upon the analysis in Sec.~\ref{sec:scenario3}, shifting the focus to robustness against reverberation. We examine how different metrics vary with the reverberation time applied to the measurements.
Figure \ref{fig:results4}a shows the $L^1$ parameter estimation error as a function of the reverberation time, \( T_{60} \). The Proposed and End-to-end methods perform similarly, with a slight increase in error at longer \( T_{60} \) values.

Figure \ref{fig:results4}b depicts the audio-domain LSD metric with respect to \( T_{60} \), where the equalization is applied to signals enhanced by DeepFilterNet2. As in Sec.~\ref{sec:scenario3}, the metrics are heavily influenced by the performance of the speech enhancement model, overshadowing the underlying robustness of the equalization methods.
To address this limitation, we repeat the experiment under an idealized scenario where perfect dereverberation is assumed. Specifically, the reference dry signal is equalized using the inverse filter estimated from the reverberant measurement. The result, presented in Fig.~\ref{fig:results4}c, shows that the correlation between the metrics and \( T_{60} \) becomes negligible under these conditions. 
In this idealized setup, the Proposed method outperforms End-to-end and demonstrates a marginal but consistent advantage over the Average method in terms of LSD, further highlighting its robustness to reverberation.

%% file: subjective_eval.tex
The goal of the listening test was to evaluate the performance of the proposed method for music equalization, focusing on subjective preference. As music equalization is inherently subjective, we aimed to determine which type of equalization participants preferred. The primary question for the test was simple: ``Which one do you prefer?''. Participants were encouraged to choose randomly when they were unsure.

The test followed a pairwise comparison preference method, where participants were presented with two audio samples and asked to choose the one they preferred. This allowed for direct comparisons between different equalization conditions. We selected four music excerpts, each approximately 7 s long, from four different genres: \textit{Jazz}, \textit{Opera}, \textit{Pop}, and \textit{Rock}. These samples are available in the attached material.

Participants assessed five distinct scenarios: the \textit{Reference} condition, featuring the untouched audio; the \textit{Distorted} condition, where the original audio sample was pre-processed with a random equalization filter before serving as input for subsequent methods; the \textit{Proposed} condition, which was treated using the technique suggested in this study; the \textit{Oracle} condition, which utilized the same equalization approach as the \textit{Proposed} but with access to the reference for calculating the average power spectra; and the \textit{Average} condition, which employed an average equalization approach based on the spectral average of the music training dataset. All samples were loudness-normalized preceding the tests to ensure that evaluations focused on spectral coloration rather than differences in loudness. Every pairwise comparison was conducted twice to eliminate potential biases that might occur if users tend to favor the letters A or B when uncertain. Additionally, all pages were displayed in a random sequence.

We implemented the test using the WebMUSHRA tool \cite{schoeffler2018webmushra}, with an unofficial version adapted specifically for preference tests \footnote{\href{https://github.com/Simon-Stone/webMUSHRA/tree/preference_test}{https://github.com/Simon-Stone/webMUSHRA/tree/preference\_test}}. 
%Figure \ref{fig:GUI} presents a screenshot illustrating the user interface used in the listening test. 
The test involved 8 participants, who all completed it in less than 15 min. The test was intentionally kept brief to prevent participant fatigue.
All participants had prior experience with listening tests. The average age was 28, with an equal gender distribution. No participants were excluded.

%\begin{figure}[t]
%    \centering
%    \includegraphics[width=0.9\linewidth]{figures/GUI.png}
%    \caption{User interface from the listening test, adapted from WebMUSHRA.}
%    \label{fig:GUI}
%\end{figure}

\subsection{Listening Test Results}

\begin{figure}[t]
\centering
    \includegraphics[trim={0 8cm 0 7cm},clip,width=0.75\linewidth]{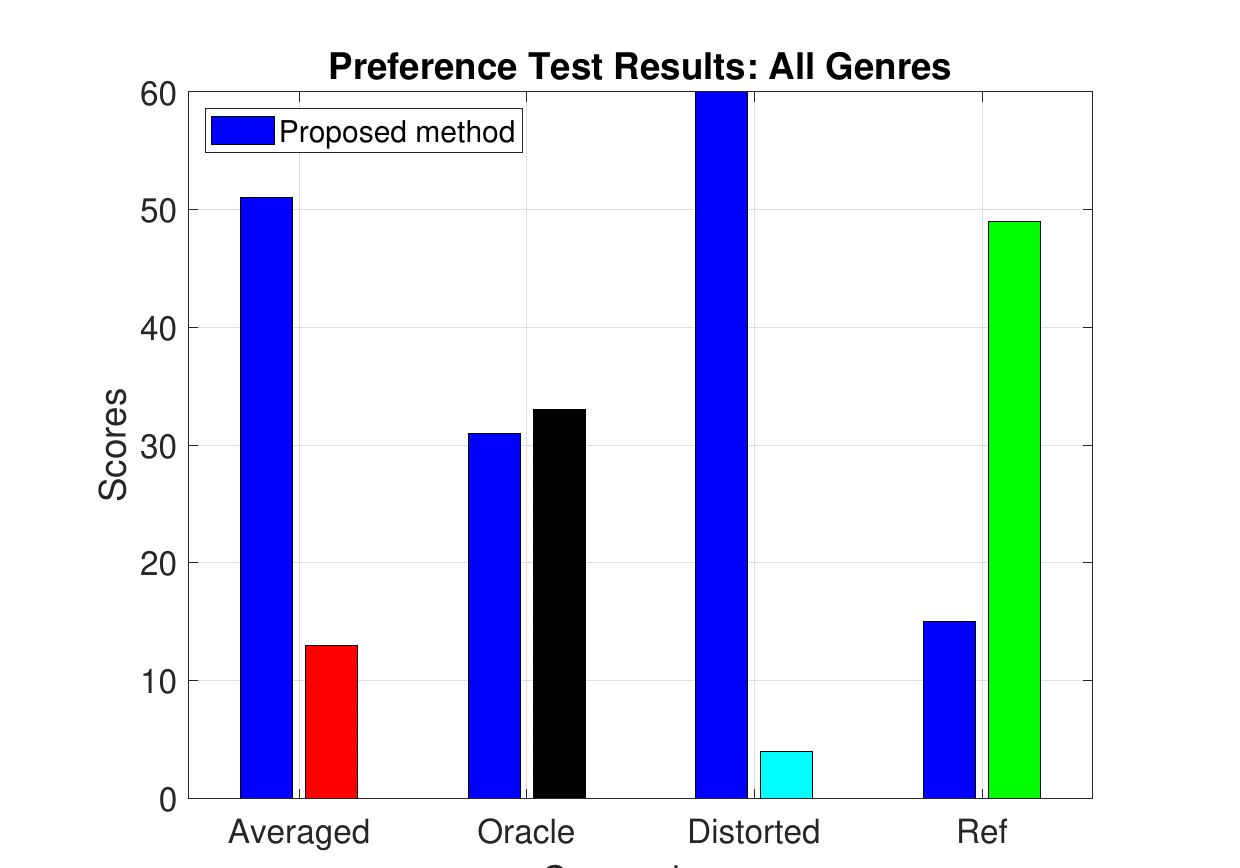}
    \captionsetup{font=small}
    \vspace{-0.2cm}
    \caption{ Listening test results from pairwise comparison of the proposed method. The data correspond to the Wins A and B in the four first lines of Table \ref{tab:all}. }
    \vspace{-0.2cm}
    \label{fig:results}
\end{figure}

The counts are aggregated across all participants and trials.
The combined results for all genres are provided in Table \ref{tab:all}.  Figure \ref{fig:results}(e) also summarizes the aggregated results of the pairwise comparison of the proposed method with the other conditions. 
 The tables show, for each possible pair, the number of wins, i.e., the times a participant chose one over the other. These tables illustrate, for each possible pair of conditions, the number of times participants preferred one option over the other.
The results from the \textit{Jazz}, \textit{Opera}, \textit{Pop} and \textit{Rock} examples are represented in Table \ref{tab:genres}.
 Figures \ref{fig:results}(a--d) also show the results of the comparison of the proposed method with the other four conditions.

For each pair, a statistical analysis was performed to determine whether the observed preference was significant. Specifically, a Z-test for proportions was conducted, comparing the proportion of times each condition was selected. The Z-statistic quantifies the difference between these proportions, taking into account the sample size and variability.
A positive Z-statistic indicates that the first condition (e.g., ``Option A") was preferred more often than the second (e.g., ``Option B"), while a negative Z-statistic suggests the opposite.
The p-value associated with the Z-statistic represents the probability that the observed difference in preferences could occur due to random chance. A p-value below 0.01 is typically considered statistically significant, suggesting that participants had a genuine preference rather than random variability. 
The tables use color coding to highlight statistically significant results. The condition preferred more frequently in significant comparisons is highlighted in green as the winner, while the less-preferred condition is highlighted in red.
\input{aggregated}

\input{genres}

%\begin{figure*}[t]
%    \centering
%    \begin{subfigure}[b]{0.23\textwidth}
%    \includegraphics[trim={0 8cm 0 7cm},clip, width=\linewidth]{figures/results_jazz.pdf}
%    \caption{Jazz}
%    \end{subfigure}
%    \begin{subfigure}[b]{0.23\textwidth}
%    \includegraphics[trim={0 8cm 0 7cm},clip,width=\linewidth]{figures/results_opera.pdf} 
%    \caption{Opera}
%    \end{subfigure}
%    \begin{subfigure}[b]{0.23\textwidth}
%    \includegraphics[trim={0 8cm 0 7cm},clip,width=\linewidth]{figures/results_pop.pdf}
%    \caption{Pop}
%    \end{subfigure}
%    \begin{subfigure}[b]{0.23\textwidth}
%    \includegraphics[trim={0 8cm 0 7cm},clip,width=\linewidth]{figures/results_rock.pdf}
%    \caption{Rock}
%    \end{subfigure}
%    
%    \begin{subfigure}[b]{0.23\textwidth}
%    \includegraphics[trim={0 8cm 0 7cm},clip,width=\linewidth]{figures/results_aggregated.pdf}
%    \caption{All genres}
%    \end{subfigure}
%    \caption{Listening test results from pairwise comparison of the proposed method with the four other conditions for the different music genres: (a) jazz, (b) opera, (c) pop, (d) rock, and (e) aggrerated. The data correspond to the Wins A and B in the four first lines of Tables \ref{tab:jazz}, \ref{tab:opera}, \ref{tab:pop}, \ref{tab:rock}, and \ref{tab:all}. }
%    \label{fig:results}
%\end{figure*}

\subsection{Discussion}

The results provide meaningful insights into the performance of the tested methods across all genres. As expected, Table \ref{tab:all} shows that the distorted condition was always rated the lowest, underscoring the detrimental impact of the applied random EQ filter on perceived audio quality.  
The proposed method outperformed or performed comparably to the average condition, with \textit{Jazz} being an exception where the proposed method and the average condition yielded similar results, as can be seen in Table \ref{tab:genres}. This highlights the capability of the proposed method to retrieve data patterns.

When compared to the \textit{Oracle} condition, the proposed method demonstrated comparable results in the aggregated analysis of Table \ref{tab:all}, with no significant differences observed. This is particularly noteworthy because it suggests that the blind power spectrum estimator employed in the proposed method nearly reaches the performance bound under the given conditions. Interestingly, in the \textit{Opera} genre, the proposed method outperformed the Oracle condition, see Table \ref{tab:genres}. This outcome may stem from the fact that the median estimate of the average power spectrum (see Eq.~\eqref{eq:median}), used in the proposed method, is subjectively more pleasing than the one derived directly from the reference.   

However, both the proposed method and the Oracle condition were consistently rated lower than the Reference condition, as seen in Table \ref{tab:all}. This points to a limitation of the inverse filter approximation. The \textit{Reference} condition serves as an idealized target, and the observed quality gap between the Oracle method and the Reference condition suggests that further refinement is needed. One promising avenue for improvement could involve increasing the resolution of the estimated coefficients by expanding the number of frequency bands in the set of parameters that the model estimates.

%% file: aggregated.tex
\begin{table}[t]
\captionsetup{font=small}
\caption{Aggregated preference listening test results across all genres. In the statistically significant cases (P $<$ 0.01), the winning and losing methods are indicated with green and red font, respectively.}
    \vspace{-0.1cm}

\label{tab:all}
\centering
\adjustbox{width=\linewidth}{
\begin{tabular}{@{}llllll@{}}
\toprule
Option A     & Option B     & Wins A            & Wins B            & Z-Statistic         & P-Value                \\ \midrule
\textcolor{green!50!black}{proposed}    & \textcolor{red!50!black}{average}      & \textcolor{green!50!black}{51}                & \textcolor{red!50!black}{13} & 6.72  & 1.85e-11         \\
proposed    & oracle       & 31                & 33               & -0.35                & 0.72                           \\
\textcolor{green!50!black}{proposed}    & \textcolor{red!50!black}{distorted}     & \textcolor{green!50!black}{60}                 & \textcolor{red!50!black}{4} & 9.90  & 4.18e-23         \\
\textcolor{red!50!black}{proposed}    & \textcolor{green!50!black}{reference}    & \textcolor{red!50!black}{15}                & \textcolor{green!50!black}{49} & -6.01  & 1.85e-09         \\ \midrule
\textcolor{green!50!black}{oracle}       & \textcolor{red!50!black}{average}      & \textcolor{green!50!black}{40}                & \textcolor{red!50!black}{24}                & 2.83  & 0.0047           \\
\textcolor{green!50!black}{oracle}       & \textcolor{red!50!black}{distorted}     & \textcolor{green!50!black}{58}                & \textcolor{red!50!black}{6}                 & 9.19   & 3.84e-20         \\
\textcolor{red!50!black}{oracle}       & \textcolor{green!50!black}{reference}    & \textcolor{red!50!black}{15}                & \textcolor{green!50!black}{49} & -6.01  & 1.85e-09         \\
\textcolor{green!50!black}{average}      & \textcolor{red!50!black}{distorted}     & \textcolor{green!50!black}{55}                & \textcolor{red!50!black}{9}  & 8.13   & 4.23e-16         \\
\textcolor{red!50!black}{average}      & \textcolor{green!50!black}{reference}    & \textcolor{red!50!black}{8}                 & \textcolor{green!50!black}{56} & -8.49  & 2.15e-17         \\
\textcolor{red!50!black}{distorted}     & \textcolor{green!50!black}{reference}    & \textcolor{red!50!black}{6}                 & \textcolor{green!50!black}{58} & -9.19  & 3.84e-20         \\
\bottomrule
\end{tabular}%
}
\end{table}

%% file: genres.tex
\begin{table}[t]
    \captionsetup{font=small}
\caption{
Results of the preference test across various genres. For brevity, paired comparisons between conditions other than the proposed method are omitted.
}
    \vspace{-0.2cm}
\label{tab:genres}
\centering
\adjustbox{width=\linewidth}{
\begin{tabular}{@{}llllllll@{}}
\toprule
Genre & Option A     & Option B     & Wins A & Wins B & Z-Statistic         & P-Value                \\ \midrule
%\textbf{\textit{Jazz}:}      &     & &  &         &                \\ 
\textbf{\textit{Jazz}}& proposed    & average       & 8                 & 8                   & 0.00                & 1.00                          \\
 & proposed       & oracle    & 8                 & 8                   & 0.00                & 1.00                          \\
& \textcolor{green!50!black}{proposed} & \textcolor{red!50!black}{distorted} &  \textcolor{green!50!black}{16}    & \textcolor{red!50!black}{0}                & 5.66   & 1.54e-08           \\
& {proposed}    & {reference}   & {5}                 & {11} & -2.12   & 0.03               \\ \midrule
%\textbf{\textit{Opera}:}      &     & &  &         &                \\ 
\textbf{\textit{Opera}}& \textcolor{green!50!black}{proposed}  & \textcolor{red!50!black}{average}    & \textcolor{green!50!black}{13}      & \textcolor{red!50!black}{3}   & 3.54   & 0.0004            \\
& \textcolor{green!50!black}{proposed}  & \textcolor{red!50!black}{oracle} & \textcolor{green!50!black}{12}      & \textcolor{red!50!black}{4}   & 2.83   & 0.0047            \\
& \textcolor{green!50!black}{proposed}  & \textcolor{red!50!black}{distorted}   & \textcolor{green!50!black}{16}      & \textcolor{red!50!black}{0}   & 5.66   & 1.54e-08          \\
& {proposed}  & {reference}    & {7}      & {9}    & -0.71   & 0.48              \\ \midrule
%\textbf{\textit{Pop}:}      &     & &  &         &                \\ 
\textbf{\textit{Pop}}& \textcolor{green!50!black}{proposed}  & \textcolor{red!50!black}{average}    & \textcolor{green!50!black}{14}      & \textcolor{red!50!black}{2}   & 4.24   & 2.21e-05  \\
& {proposed}  & {oracle} & {7}      & {9}    & -0.71    & 0.48      \\
& \textcolor{green!50!black}{proposed}  & \textcolor{red!50!black}{distorted}   & \textcolor{green!50!black}{15}      & \textcolor{red!50!black}{1}   & 4.95   & 7.43e-07  \\
& \textcolor{red!50!black}{proposed}  & \textcolor{green!50!black}{reference}    & \textcolor{red!50!black}{2}      & \textcolor{green!50!black}{14}   & -4.24   & 2.21e-05  \\ \midrule
%& \textbf{\textit{Rock}:}      &     & &  &         &                \\ 
\textbf{\textit{Rock}}& \textcolor{green!50!black}{proposed}  & \textcolor{red!50!black}{average}    & \textcolor{green!50!black}{16}      & \textcolor{red!50!black}{0}   & 5.66   & 1.54e-08  \\
& \textcolor{red!50!black}{proposed}  & \textcolor{green!50!black}{oracle} & \textcolor{red!50!black}{4}     & \textcolor{green!50!black}{12}    & -2.83    & 0.0047    \\
& \textcolor{green!50!black}{proposed}  & \textcolor{red!50!black}{distorted}   & \textcolor{green!50!black}{13}      & \textcolor{red!50!black}{3}   & 3.54   & 0.0004    \\
& \textcolor{green!50!black}{proposed}  & \textcolor{red!50!black}{reference}    & \textcolor{green!50!black}{1}      & \textcolor{red!50!black}{15}   & -4.95   & 7.43e-07   \\  \bottomrule
\end{tabular}%
}
\end{table}